\documentclass[aps,prl,preprint,superscriptaddress,amsmath,amssymb]{revtex4}

\usepackage{graphicx} 
\usepackage{dcolumn}  
\usepackage{bm}    
\usepackage{amssymb}  

\begin{document}

\title{Topological Surface States and Superconductivity in [Tl$_4$](Tl$_{1-x}$Sn$_{x}$)Te$_3$ Perovskites}

\author{K. E. Arpino}
\thanks{These authors contributed equally to this work.}
\author{D. C. Wallace}
\thanks{These authors contributed equally to this work.}
\affiliation{Department of Chemistry, The Johns Hopkins University, Baltimore, MD 21218, USA}
\affiliation{Institute for Quantum Matter, Department of Physics and Astronomy, The Johns Hopkins University, Baltimore, MD 21218, USA}

\author{Y. F. Nie}
\thanks{These authors contributed equally to this work.}
\affiliation{Department of Physics, Laboratory of Atomic and Solid State Physics, Cornell University, Ithaca New York 14853, USA}

\author{T. Birol}
\affiliation{School of Applied and Engineering Physics, Cornell University, Ithaca, NY 14853, USA}

\author{P. D. C. King}
\affiliation{Department of Physics, Laboratory of Atomic and Solid State Physics, Cornell University, Ithaca New York 14853, USA}
\affiliation{Kavli Institute at Cornell for Nanoscale Science, Cornell University, Ithaca, NY 14853, USA}

\author{S. Chatterjee}
\author{M. Uchida}
\affiliation{Department of Physics, Laboratory of Atomic and Solid State Physics, Cornell University, Ithaca New York 14853, USA}

\author{S. M. Koohpayeh}
\author{J.-J. Wen}
\affiliation{Institute for Quantum Matter, Department of Physics and Astronomy, The Johns Hopkins University, Baltimore, MD 21218, USA}

\author{C. J. Fennie}
\affiliation{School of Applied and Engineering Physics, Cornell University, Ithaca, NY 14853, USA}

\author{K. M. Shen}
\affiliation{Department of Physics, Laboratory of Atomic and Solid State Physics, Cornell University, Ithaca New York 14853, USA}
\affiliation{Kavli Institute at Cornell for Nanoscale Science, Cornell University, Ithaca, NY 14853, USA}

\author{T. M. McQueen}
\email{mcqueen@jhu.edu}
\affiliation{Department of Chemistry, The Johns Hopkins University, Baltimore, MD 21218, USA}
\affiliation{Institute for Quantum Matter, Department of Physics and Astronomy, The Johns Hopkins University, Baltimore, MD 21218, USA}

\date{\today}

\begin{abstract}
{\bf Materials with strong spin-orbit coupling have attracted attention following the prediction and subsequent discovery of strong two- and three-dimensional topological insulators in which a topological property of the bulk band structure of an insulator results in metallic surface states with Dirac-like dispersion. Here we report the discovery of Dirac-like surface states in the perovskite superconductor [Tl$_4$]TlTe$_3$ (Tl$_5$Te$_3$) and its non-superconducting tin-doped derivative, [Tl$_4$](Tl$_{0.4}$Sn$_{0.6}$)Te$_3$, as observed by angle-resolved photoemission spectroscopy (ARPES). Density functional theory (DFT) calculations predict a single spin-orbit driven band parity inversion at the $Z$ point above the Fermi level of Tl$_5$Te$_3$, suggesting the surface states are protected by Z$_2$ topology. Calculations on [Tl$_4$]SnTe$_3$ show no parity inversions, implying that a topological transition from non-trivial to trivial must occur upon doping with tin, i.e., [Tl$_4$](Tl$_{1-x}$Sn$_{x}$)Te$_3$. Thus [Tl$_4$]{\it M}Te$_3$ perovskites are a possible new, non-trigonal class of Z$_2$ topological compounds. Additionally, as Tl$_5$Te$_3$ is a stoichiometric bulk superconductor, these perovskites are ideal materials in which to study the interplay between surface states and bulk superconductivity.}
\end{abstract}
\maketitle

Charge carriers in Dirac-like bands have a linear energy-momentum relationship described by the Dirac equation; they offer the unique opportunity to investigate the intersection of special relativity and quantum mechanics. However, materials exhibiting Dirac surface states are relatively rare, the most notable examples being graphene \cite{Graphene}, Z$_2$ topological insulators \cite{TIpredict1}, and topological crystalline insulators (TCIs) \cite{TCIpredictFu}. Experimentally known topological insulators are based on a limited range of materials and structure types: strained HgTe, Bi$_x$Sb$_{1-x}$, (Bi/Sb)$_2$(Se/Te)$_3$, Tl[Bi/Sb](S/Se/Te)$_2$, and Bi$_{14}$Rh$_3$I$_9$ \cite{HgTeQW,SbdopedBi,Bi2X3TI,TlBiSe2Class,Bi14Rh3I9} with many others theoretically predicted \cite{PbTI,LaPtBi,FelserPredict,ColemanKondoTI}. Materials which combine topological surface states with superconductivity have the ability to test the predictions of new states of matter and exotic quasiparticles, such as Majorana fermions, and open the field for new technologies such as quantum computing \cite{app1,app2,app3,app4}. Producing such materials has been a challenge: the known methods of combining these two properties are proximity effects and doping (Cu-doping of Bi$_2$Se$_3$ and In-doping of Sn$_{1-x}$In$_x$Te)\cite{ProximitySCTI,CuSCTI,SCTCI}, which limits studies to small samples and/or materials with low volume fraction superconductivity. 

Here we report the discovery of Dirac-like surface states in the tetragonal perovskites [Tl$_4$](Tl$_{1-x}$Sn$_{x}$)Te$_3$, as observed by angle-resolved photoemission spectroscopy (ARPES). The surface states are predicted by density functional theory (DFT) to be Z$_2$ topologically protected due to a spin-orbit-driven band parity inversion at the $Z$ point. This is the first observation of Dirac-like surface states in a material with the perovskite structure and represents a possible new, non-trigonal class of Z$_2$ topological compounds. Moreover, we show that $\sim$100\% volume fraction superconductivity is possible in Tl$_5$Te$_3$, making it an ideal compound in which to explore the coupling between bulk superconductivity and Dirac-like surface states, i.e., a ``topological superconductor."

\section{Results}
\subsection{Angle-Resolved Photoemission Spectroscopy}
\begin{figure}
\includegraphics[width=3.5in]{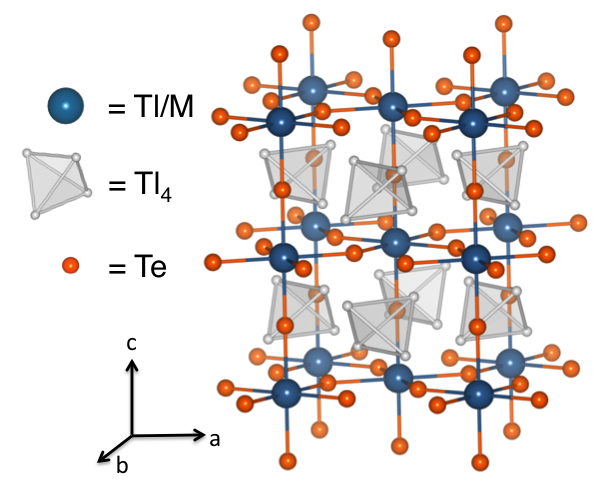}
\caption{(Color online) {\bf Structural model of the perovskite framework of  [Tl$_4$]$M$Te$_3$ (M = metal).} Tl$_4$ tetrahedra occupy the $A$ site of a conventional perovskite $AB$O$_3$, and the alternating rotation of $M$Te$_6$ octahedra about the {\it c} axis is described as  $a^0$$a^0$$c^-$ in Glazer notation with a rotation angle of 20.1$^{\circ}$ for Tl$_5$Te$_3$ \cite{Glazer,Struct1,Vesta}.}
\label{figstruct}
\end{figure}

The [Tl$_4$]{\it M}Te$_3$ ($M$ = Sn, Pb, Bi, Sb, La, Nd, and Mo \cite{Dop4,DopLa,DopNd,DopMo}) family of compounds has perovskite structure ($AB$O$_3$) described by Glazer notation $a^0$$a^0$$c^-$ with body-centered tetragonal symmetry \cite{Struct1,Glazer}. The cavities of a three-dimensional network of corner-sharing $M$Te$_6$ octahedra are occupied by interconnected Tl$_4$ tetrahedra [Fig.~\ref{figstruct}(a)].

In Fig.~\ref{figSnTI}(a) we show ARPES data on [Tl$_{4}$](Tl$_{0.4}$Sn$_{0.6})$Te$_{3}$. The most prominent feature is an intense, linearly dispersing, Dirac-like band centered at a momentum of (k$_{x}$ = 0,k$_{y}$ = 0) 
of the two-dimensional surface Brillouin zone (2DBZ). We observe this feature at the center of numerous Brillouin zones, and only a single such metallic feature in each zone.
We assign this Dirac-like band to be a surface state based on a number of different reasons. First, the values of the extracted Fermi velocity [4.6(2) eV \AA] and the Fermi wavevector [0.04(1) \AA$^{-1}$] are insensitive to incident photon energy indicating that they are non-dispersive in k$_{z}$ [Fig.~\ref{figSnTI}(c)], even though the bulk electronic structure is highly three-dimensional. This is a hallmark of a surface state. Furthermore, this feature is present and always centered at (0,0) of the 2DBZ, and not the center of the three-dimensional bulk Brillouin zone, as would be expected from a bulk-derived state. Additionally, we also observe features near the Fermi energy ($E_{F}$) that do vary considerably with the incident photon energy, and which we ascribe to bulk states, consistent with DFT calculations (see below). 

In Fig.~\ref{figTlTI}, we show ARPES data from Tl$_{5}$Te$_{3}$. Much like in [Tl$_{4}$](Tl$_{0.4}$Sn$_{0.6})$Te$_{3}$, a single, Dirac-like surface state is once again visible at (0,0)
 while no other Dirac-like features are observed anywhere else in the 2DBZ. As Tl$_{5}$Te$_{3}$ is metallic, containing fewer electrons than [Tl$_{4}$](Tl$_{0.4}$Sn$_{0.6})$Te$_{3}$, only the bottom section of the Dirac cone is visible, and the extrapolated Dirac point sits 200 meV above E$_{F}$. The Fermi velocities [3.0(2) eV \AA] and Fermi wavevectors [0.07(1) \AA$^{-1}$] observed for the Dirac-like band in Tl$_{5}$Te$_{3}$ are largely independent of photon energy ($h\nu$ = 21.2 and 40.8 eV). 

The similarities between surface states observed both in [Tl$_{4}$](Tl$_{0.4}$Sn$_{0.6})$Te$_{3}$ and Tl$_{5}$Te$_{3}$ and the single Dirac cone observed per Brillouin zone suggest that [Tl$_4$](Tl$_{1-x}$Sn$_{x}$)Te$_3$ host topologically protected surface states. Further indirect evidence for their topological nature is that we find no Rashba-type spin splitting of the surface state despite the strong spin-orbit coupling of these materials, suggesting that we have just a single spin non-degenerate Fermi surface between neighboring time-reversal invariant points, the hallmark of a TI.

\subsection{Density Functional Theory Calculations}
However, there are several ways in which the [Tl$_4$]$M$Te$_3$ structure can result in topological Dirac-like surface states. First, the mirror planes which topologically protect the Dirac states in rock salt TCIs are also present in the basic perovskite structure. The appropriate band inversions at points protected by mirror symmetry similarly may yield a TCI state in body-centered tetragonal symmetry. Second, the body-centered tetragonal crystal class has four sets of time-reversal invariant points: one $\Gamma$, four $N$, two $X$, and one $Z$. If spin-orbit coupling produces band parity inversions at the $\Gamma$ and $Z$ points, but not $N$ or $X$, surface states protected as in a weak topological insulator result. Alternately, a band inversion due to spin-orbit coupling at the $\Gamma$ or $Z$ point, but not both, produces strong Z$_2$ topologically protected surface states. 

\begin{figure}
\includegraphics[width=5.5in]{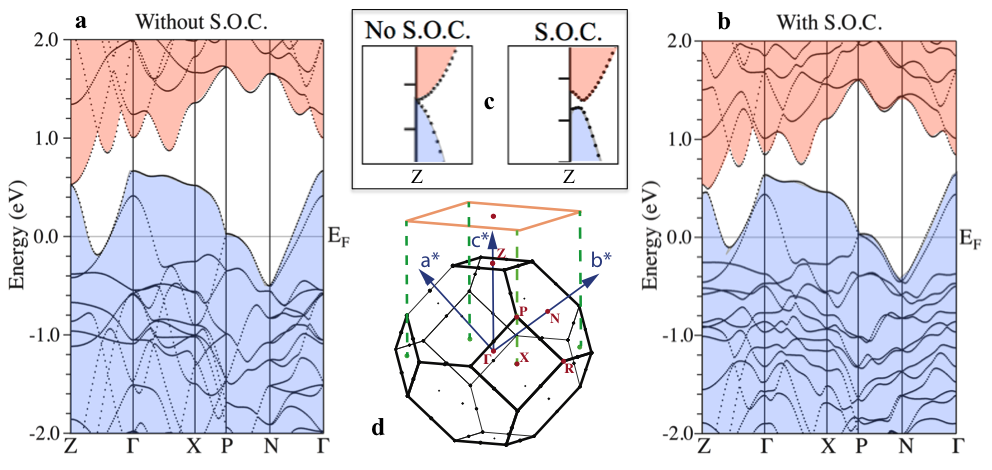}
\caption{(Color online) {\bf Spin-orbit driven band inversion in Tl$_5$Te$_3$.}The band structure of Tl$_5$Te$_3$, calculated from first principles DFT, {\bf (a)} without and with {\bf (b)} spin-orbit interactions. Shading denotes bands arising from valence (light blue) and conduction (light red) band states. {\bf (c)} The spin-orbit interaction produces a single band parity inversion and opens a small gap at the $Z$ point, indicating that Tl$_5$Te$_3$ would be a topological semi-metal if doped to the gap. {\bf (d)} The Brillouin zone of the body-centered tetragonal unit cell for comparison with the ARPES results.}
\label{figDFTTl}
\end{figure}

Calculations for Tl$_5$Te$_3$, shown in Fig.~\ref{figDFTTl}, show spin-orbit coupling induces
a single band inversion approximately 0.5 eV above the Fermi level between valence and conduction band states at the $Z$ point. This inversion is between opposite-parity states, the conduction band involving states from the [Tl$_4$] units on the $A$ site and the valence band involving mainly states from apical Te atoms and $B$ site Tl ions in the corner-sharing perovskite network. If electron-doped to the gap at $Z$ opened by spin-orbit coupling, this system would be a topological semi-metal due to high-lying valence band states at the $\Gamma$ point. The resulting situation is thus similar to that found in pure antimony \cite{SbdopedBi}. Furthermore, we have calculated the Z$_2$ invariant to be -1 by taking product of the parity of all the bands at the $\Gamma$ and $Z$ points that connect to the occupied valence bands, including two bands above the Fermi level at the $\Gamma$ point. Therefore, DFT predicts Tl$_5$Te$_3$ to have surface states protected by a non-trivial Z$_2$ topology. 

\begin{figure}
\includegraphics[width=5.5in]{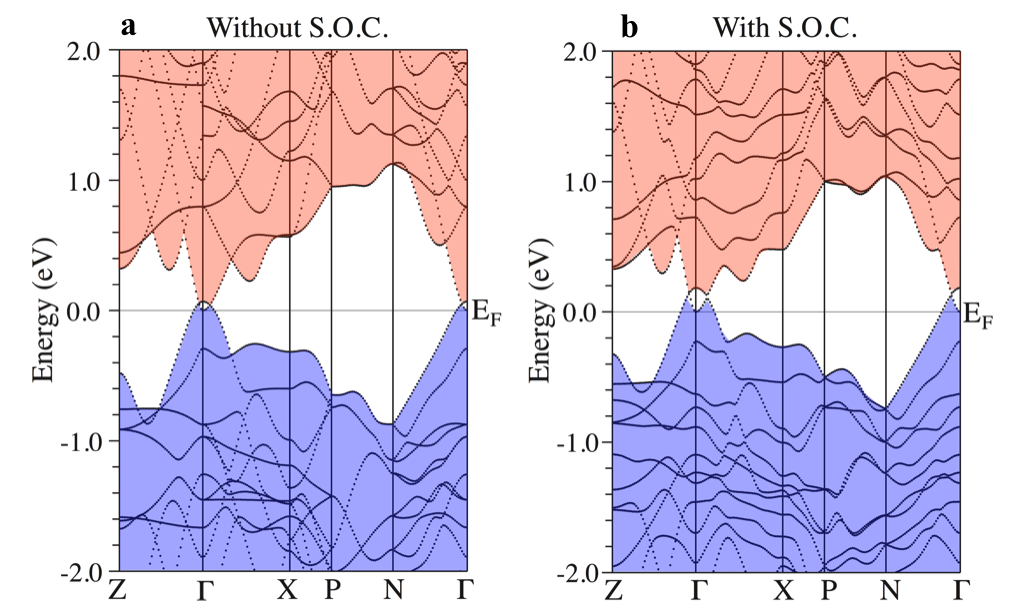}
\caption{(Color online) {\bf Topologically trivial nature of [Tl$_4$]SnTe$_3$.} The band structure of Tl$_4$SnTe$_3$, calculated from first principles DFT, is shown {\bf (a)} without and with {\bf (b)} spin-orbit interactions. Shading denotes bands arising from valence (light blue) and conduction (light red) band states. Unlike the Tl$_5$Te$_3$ case, the spin-orbit interaction does not produce a band parity inversion at the $Z$ point, implying that [Tl$_4$]SnTe$_3$ is topologically trivial semi-metal. The band inversion near the $\Gamma$ point is between bands of the same parity and does not affect topological class.}
\label{figDFTSn}
\end{figure}

In contrast, our calculations shown in Fig.~\ref{figDFTSn} demonstrate that [Tl$_4$]SnTe$_3$, the opposite end member of the [Tl$_4$](Tl$_{1-x}$Sn$_{x}$)Te$_3$ solid solution, has trivial Z$_2$ topology: although a possible band inversion occurs near the $\Gamma$ point, the involved states are of the same parity. Further, an approximately 0.7 eV band gap opens up at the $Z$ point where the band crossing is observed in Tl$_5$Te$_3$, preventing spin-orbit coupling from inverting band parity at that point. These calculations imply that a transition from Z$_2$ non-trivial  (at $x$ =0, or Tl$_5$Te$_3$) to trivial (at $x$ =1, or [Tl$_4$]SnTe$_3$) topology must exist at some doping level $x$ in [Tl$_4$](Tl$_{1-x}$Sn$_{x}$)Te$_3$ compounds. The situation in the [Tl$_4$](Tl$_{1-x}$Sn$_{x}$)Te$_3$ peroskites is thus similar to what is found for the Bi$_{1-x}$Sb$_{x}$ solid solution \cite{SbdopedBi}. 

To understand the driving force behind the topological transition, we investigated the origins of the large gap opening at the $Z$ point when Sn is substituted for Tl. There are two major structural changes when this doping is performed: a sizable change in lattice parameters, particularly the increase in the $c$ axis relative to $a$, and cation mixing. In order to isolate the effect of lattice parameters, the band structure of Tl$_5$Te$_3$ was calculated with the same lattice constants as [Tl$_4$]SnTe$_3$ imposed, and it was found that Tl$_5$Te$_3$ would then be topologically trivial, albeit with a smaller bandgap at the $Z$ point. Further emphasizing the importance of the lattice parameters on the topological transition is the fact that the width of the valence band involved in the band inversion that has large apical Te weight depends sensitively on the lattice constant $c$ of the material due to slight $B$ site Tl contribution. This therefore demonstrates the importance of lattice parameters in determining the Z$_2$ topological invariant in the [Tl$_4$](Tl$_{1-x}$Sn$_{x}$)Te$_3$ compounds. Preliminary calculations (not shown) suggest that the effect of cation ordering is also significant, but detailed studies beyond the scope of this work are needed to elucidate those details and to determine the critical value of $x$ required for the transition to a different topological phase in [Tl$_4$](Tl$_{1-x}$Sn$_{x}$)Te$_3$ compounds. The sensitivity to structural parameters, particularly the $c/a$ ratio, suggest that temperature- and/or strain-driven topological transitions may be possible in this material family. Regardless of these details, DFT results are consistent with the ARPES detection of a Dirac cone in both the Tl$_5$Te$_3$ and [Tl$_4$](Tl$_{0.4}$Sn$_{0.6}$)Te$_3$ compounds and indicate that Z$_2$ topology protects surface states in this materials family.

\subsection{Superconducting Properties}
\begin{figure}
\includegraphics[width=3.5in]{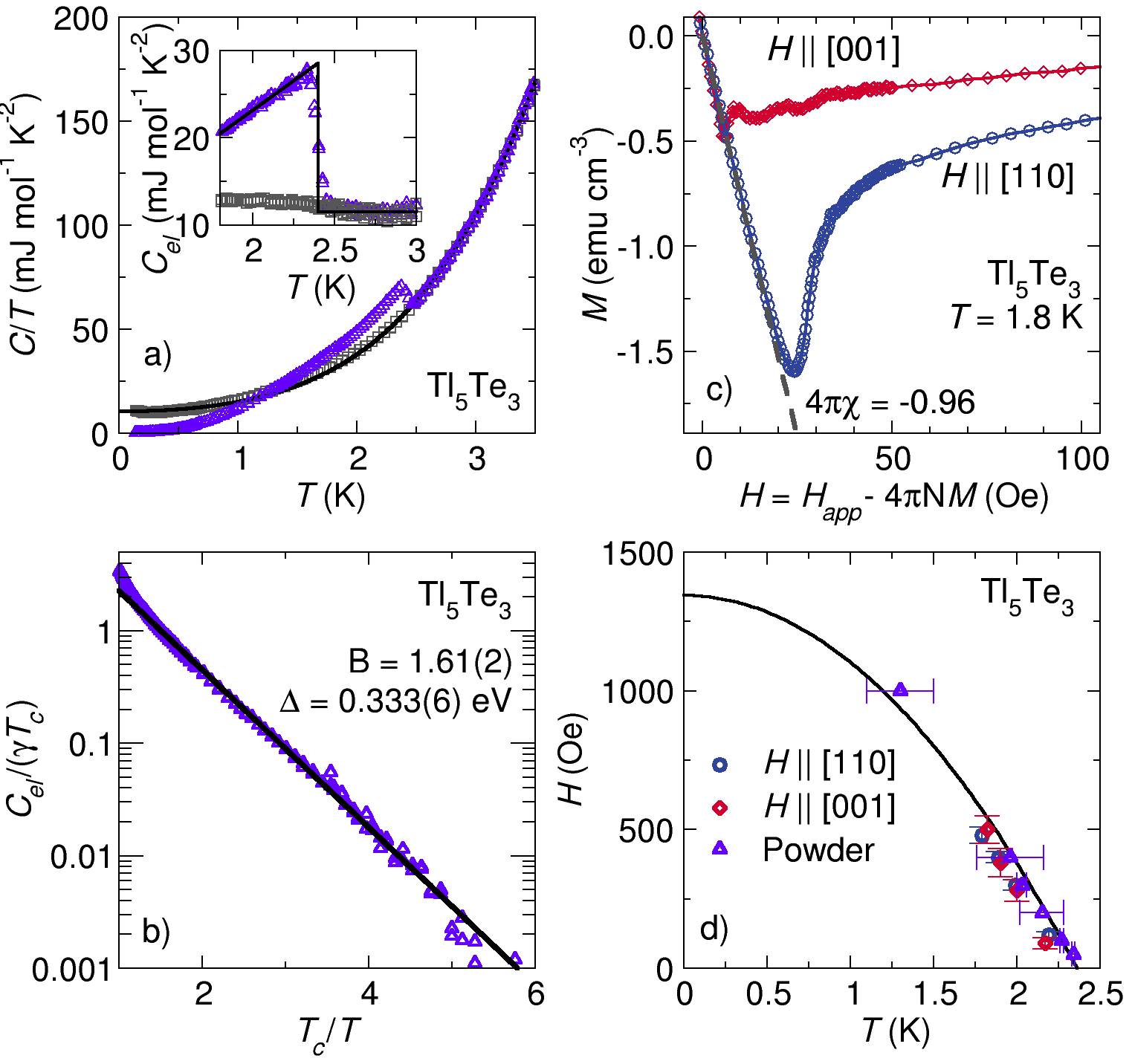}
\caption{(Color online) {\bf Superconducting properties of topological Tl$_5$Te$_3$.} {\bf (a)} Specific heat data collected under zero field (purple triangles) show a clear $\lambda$-anomaly at $T_c$ = 2.40(1) K, while those collected under an applied 5 kOe  field (gray squares) fit well to $C_n$ = $\gamma T$ + $\beta _3$$T$$^{3}$ + $\beta _5$$T$$^{5}$.  Inset: The electronic specific heat as a function of temperature is shown along with an equal entropy construction illustrating bulk superconductivity with $T_c$ = 2.40(1) K. {\bf (b)} Low-temperature electronic specific heat behaves in accordance with single-gap BCS theory. {\bf (c)} Isothermal magnetization data collected on an oriented single crystal at $T$ = 1.8 K with the field applied parallel to the [001] direction (red diamonds) and [110] direction (blue circles) reveal a striking anisotropy of $H_c^*$. The slope of the data in the linear region below $H_c^*$ indicate a 96\% superconducting volume fraction. A demagnetization factor N = 1/3 has been applied to account for the approximately cubic shape of the 100 mg single crystal sample. {\bf (d)} The field-temperature phase diagram shows $H_{c2}$ as determined by specific heat (purple triangles) and orientation-dependent isothermal DC magnetization (blue circles and red diamonds), along with a two-fluid model fit to the data. The orientation anisotropy present in $H_{c,i}^*$ is not observed for $H_{c2}$.}
\label{figsc}
\end{figure}

The observation of Dirac-like surface states in Tl$_5$Te$_3$, likely Z$_2$ topologically protected, makes it a candidate topological superconductor. Superconductivity was not observed in [Tl$_{4}$](Tl$_{0.4}$Sn$_{0.6})$Te$_{3}$ above $T$ = 1.8 K. Because Tl$_5$Te$_3$ was previously proposed to be a two-band superconductor \cite{JK1968,JK1973}, we performed a comprehensive specific heat and magnetization study of the superconducting properties. Our specific heat data collected on a polycrystalline chunk under zero magnetic field reveal a clear $\lambda$-anomaly. An equal-entropy construction, shown in Fig.~\ref{figsc}(a), indicates bulk superconductivity with a $T_c$ = 2.40(1) K. The ratio $\Delta C_{el}$/($\gamma T_c$) = 1.63 is close to, but slightly greater, than the weak coupling Bardeen-Cooper-Schrieffer (BCS) value of 1.43 \cite{BCS}. The normal state heat capacity, measured under {$H$} = 5 kOe, is well described by $C_n$ = $\gamma T$ + $\beta _3$$T$$^{3}$ + $\beta _5$$T$$^{5}$,  where the term linear in $T$ represents the electronic contribution and the higher order terms represent the lattice contribution \cite{SpecificHeat}. The fit yields an electronic Sommerfeld coefficient of $\gamma$ = 10.6(1) mJ mol f.u.$^{-1}$ K$^{-2}$. The coefficients $\beta_3$ and $\beta_5$ of the lattice contribution are 3.90(5) mJ mol f.u.$^{-1}$ K$^{-4}$ and 0.740(6) mJ mol f.u.$^{-1}$ K$^{-6}$, respectively, and the $\beta_3$ corresponds to a Debye temperature of $\Theta _D$ = 159(2) K \cite{SpecificHeat}.

Low-temperature specific heat measurements collected under $H$ = 0 Oe and 5 kOe show the latter field is sufficient to quench superconductivity. The data collected under a $H$ = 5 kOe field below $T$ = 1.5 K, temperatures at which any complex phonon structure should be negligible, were fit to $C_{el}$ = $\gamma$$T$ + $\beta _3$$T^{3}$, where $\gamma$ = 9.34(4)  mJ mol f.u.$^{-1}$ K$^{-2}$ and $\beta _3$ = 6.82(5)  mJ mol f.u.$^{-1}$ K$^{-4}$ \cite{SpecificHeat}. Subtracting this low-temperature lattice specific heat fit gives the electronic specific heat, which is well described by the BCS model of the electronic specific heat of a single-gap $s$-wave superconductor $C_{el}$/($\gamma$$T_c$) = Aexp[(-B$T_c$)/$T$] with A =  11.4(1.1) and B = 1.61(2) [Fig.~\ref{figsc}(b)] \cite{SpecificHeat, BCS}. According to BCS theory, B = $\Delta$ $k_\text{B}$$ T_c$, and thus a value of  B = 1.61(2) predicts a superconducting gap of $\Delta$ = 0.333(6) meV. This value, based on the electronic specific heat data, agrees with the superconducting gap of $\Delta$ = 0.365(2) meV predicted using the critical temperature of $T_c$ = 2.40(1) K in the BCS relation 2$\Delta$/($k_\text{B}$$T_c$) = 3.53. 

Isothermal DC magnetization measurements were collected on an approximately cubic 100 mg single crystal of Tl$_5$Te$_3$ with (001) and (110) faces, allowing the orientation dependence to be examined. At $T$ = 1.8 K, a full superconducting volume fraction is observed from both crystal directions under small applied magnetic field, and the upper critical fields, H$_{c2, ab}$(1.8 K) = 480(30) Oe and H$_{c2, c}$(1.8 K) = 500(50) Oe are within error. In contrast, the lower critical fields are found to differ by a factor of greater than three, as shown in Fig. \ref{figsc}(c):  $H^*_{ab}$(1.8 K) = 18(1) Oe and  $H^*_{c}$(1.8 K) = 5(1) Oe, where  $H^*_{i}$(1.8 K) is defined as the magnetic field, parallel to the {\it i} direction, at which the magnetic response deviates from a linear field-dependence. We call this  $H^*_{i}$ rather than the strictly defined lower critical field $H_{c1, i}$ because the presence of surface states may produce spurious effects in the measured magnetic response. Note this observed threefold anisotropy is merely a lower bound of the extent of the anisotropy present in the bulk superconducting state: any intrinsic material defect of the crystal or minor experimental misalignment would result in contribution from the perpendicular direction. In short, Tl$_5$Te$_3$  appears to be a single-gap BCS superconductor. It is thus an ideal material in which to probe whether topological surface states participate in superconductivity, or how these gapless states affect, and are affected by, the bulk superconducting gap.

\section{Discussion}
Charge carriers in Dirac-like bands have a linear energy-momentum relationship described by the Dirac equation; they offer the unique opportunity to investigate the intersection of special relativity and quantum mechanics. The Dirac cones of graphene are even-numbered and topologically trivial. In contrast, the surface states of Z$_2$ topological insulators and TCIs are protected topologically by time-reversal and mirror symmetry, respectively. Strong Z$_2$ topological insulators must have an odd number of band inversions at time-reversal invariant points to ensure protected surface states. Electron transport on the surface of a Z$_2$ topological insulator is novel because the spin and momentum of surface electrons are ``locked" together. For this reason, topological insulators have a wide variety of potential applications. 

\begin{figure}
\includegraphics[width=5.5in]{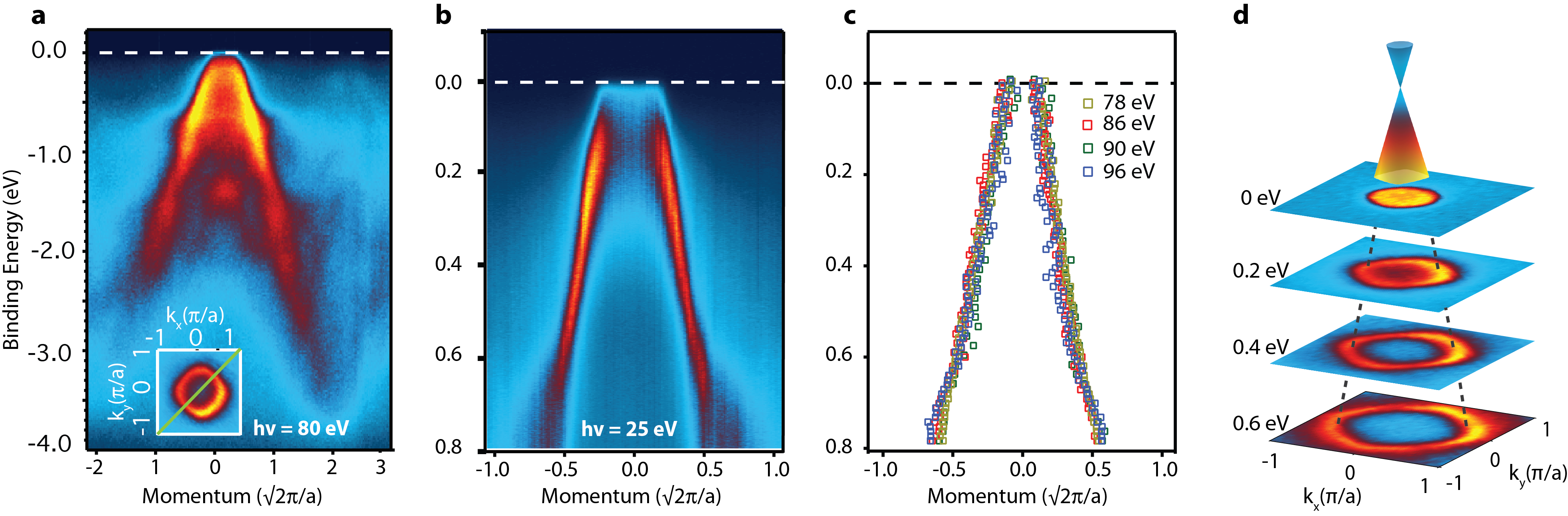}
\caption{(Color online) {\bf Dirac-like surface state in [Tl$_4$](Tl$_{0.4}$Sn$_{0.6})$Te$_3$. (a)} ARPES spectrum taken along the (0,0) -- ($\pi$,$\pi$) shows the electronic states over a wide energy range, including bulk states, and a linearly dispersing surface state which evolves into a surface resonance upon intersecting the bulk states. {\bf (b)} Expanded view of the Dirac-like surface state within the projected bulk band gap. {\bf (c)} Fits to the dispersion at various photon energies, revealing no substantial dispersion along k$_z$. {\bf (d)} Momentum-resolved intensity maps plotted at different binding energies, which suggest the bottom portion of a single Dirac cone centered at the zone center, consistent with a topological surface state, with a square warping at higher binding energies, reflecting the tetragonal crystal symmetry.}
\label{figSnTI}
\end{figure}

Our results suggest a number of characteristics which make the [Tl$_4$]$M$Te$_3$ family an intriguing alternative to most other topological insulator families. First, as shown in Fig.~\ref{figSnTI}(d), the truncation of the Dirac cone indicates that [Tl$_{4}$](Tl$_{0.4}$Sn$_{0.6})$Te$_{3}$ is $p$-type doped, in contrast to materials such as Bi$_{2}$Se$_{3}$, which are typically $n$-type in the bulk. Second, instead of the six-fold warping of the Dirac cone away from the Dirac point which is consistent with the trigonal structure of most topological insulators, the Dirac cone in [Tl$_{4}$](Tl$_{0.4}$Sn$_{0.6})$Te$_{3}$ exhibits a four-fold warping consistent with its tetragonal symmetry. At higher binding energies, the surface state appears to intersect and begin to hybridize with the bulk bands [such as around 0.6 eV in Fig.~\ref{figSnTI}(a)], forming a surface resonance with some k$_{z}$ dependence in the Fermi velocity. However, at energies near E$_{F}$ within the bulk bandgap, this band exhibits a slightly higher Fermi velocity shown in Fig.~\ref{figSnTI}(c). Third, DFT analysis shows that the topological transition from trivial to non-trivial must occur in [Tl$_{4}$](Tl$_{1-x}$Sn$_{x})$Te$_{3}$ as $x$ decreases. The sensitivity of this transition to structural parameters, particularly the $c$/$a$ ratio, suggesting that temperature-, pressure-, or stain-induced topological transition may be possible in this material family.

The coexistence of topologically protected Dirac-like surface states and superconductivity in Tl$_5$Te$_3$ make it a candidate ``topological superconductor." Topological superconductivity is highly sought after due to the prediction of massless Majorana fermions, which have great potential to advance technology, specifically in the area of quantum computing, and to improve our knowledge of fundamental physics. Although superconductivity has previously been achieved in topological systems via proximity effects and by doping, experiments investigating the phenomenon are hindered by several factors. For example, while superconductivity has previously been realized in Cu$_x$Bi$_2$Se$_3$ by intercalating copper into the Van der Waals gaps between layers, a small superconducting volume fraction is observed in this material \cite{CuSCTI}. Further, the substitution of copper for bismuth (Bi$_{2-x}$Cu$_x$Se$_3$) competes with copper intercalation (Cu$_x$Bi$_2$Se$_3$), thereby limiting the attainable Cu dopant concentration and producing inhomogeneous samples \cite{CuSCTI}. Superconductivity in topological insulators has also been successfully induced in Bi$_2$(Se/Te)$_3$ by the proximity effect, but studies are then limited to thin sample regions\cite{ProximitySCTI}. Sn$_{1-x}$In$_x$Te and LaPtBi are also candidate materials, but the former is a TCI and the latter lacks a center of inversion, complicating efforts to understand them \cite{SCTCI,LaPtBi}.

In contrast to these materials,
the [Tl$_4$]$M$Te$_3$ family provides an ideal system to investigate the effects of topologically protected surface states on superconductivity because Tl$_5$Te$_3$ is an intrinsic bulk superconductor that is straightforward to synthesize and computationally model. Tl$_5$Te$_3$'s bulk superconductivity eliminates the experimental obstacles faced in doped or proximity-induced topological superconductors, such as low superconducting volume fraction, sample inhomogeneity, and size limitations. Further, the wide range of cations reported to substitute for the $M$ site, as well as the possibility of substituting other chalcogens or halogens for tellerium, presents a versatile range of experimentally accessible doping levels and resultant properties to examine. 

The [Tl$_4$]$M$Te$_3$ family has an entirely different crystal structure than the known Z$_2$ topological insulators, which all have trigonal symmtry in the active units. Because symmetry determines the type and number of time-reversal invariant momentum points in the Brilliouin zone of a material as well as its mirror planes, the structure type of a material is an important factor in whether Dirac surface states exist and whether they are topologically non-trivial. Our finding of topological surface states in perovskites therefore has important materials implications. Other members of the [Tl$_4$]{\it M}Te$_3$ family likely exhibit similar topological properties in conjunction with other emergent phenomena including superconductivity. More generally, our results open the door to finding topologically protected states in other members of the largest known material family -- perovskites.

\section{Methods}
\subsection{Sample Preparation}
Polycrystalline Tl$_5$Te$_3$ samples were prepared by heating elemental Tl (Strem Chemicals, 99.9 $\%$, {\bf warning: toxic, carcinogenic}) and Te (Alfa Aesar, 99.999+ $\%$) in a vacuum-sealed silica ampoule to 550$^{\circ}$C, holding for 24 h, followed by slow cooling (5$^{\circ}$C/h). Elemental thallium was stored and handled in an inert atmosphere to prevent oxidization. Single crystals were prepared using a modified Bridgman method in an optical floating-zone furnace (Crystal Systems Inc.) using 2.5\% excess Te as a flux. Tl$_{4}$(Tl$_{1-x}$Sn$_{x})$Te$_{3}$ samples were prepared by heating elemental Tl, Te, and Sn (NOAH Technologies, 99.9$\%$) in a vacuum-sealed silica ampuole to 540$^{\circ}$C, holding for 24 h, followed by slow cooling (2$^{\circ}$C/h). Single crystals of Tl$_{4}$(Tl$_{0.4}$Sn$_{0.6})$Te$_{3}$ suitable for ARPES measurements were obtained from the ingot after slow cooling.

\subsection{Physical Properties Measurements}
Physical property measurements were performed on a Quantum Design, Inc. Physical Property Measurement System. Low-temperature heat capacity data from $T$ = 0.15 K to 4 K were collected on a polycrystalline chunk with the aid of a dilution refrigerator. ARPES measurements were performed at the Synchrotron Radiation Center in Wisconsin at the U3 beamline with a VG Scienta R4000 electron analyzer at a base temperature of 30 K and a pressure of 8 $\times$ 10$^{-11}$ torr as well as a lab-based ARPES system with a VG Scienta R4000 analyzer and VUV5000 helium plasma discharge lamp ($h\nu$ = 21.2 and 40.8 eV) and monochromator at a temperature of 10 K and a base pressure of 6 $\times$ 10$^{-11}$ torr. Samples were cleaved \emph{in situ} along the $c$ axis in ultra-high vacuum in order to expose atomically clean surfaces. Due to the three-dimensionally bonded structure, cleaved samples did not yield entirely flat surfaces, but still possessed regions with reasonably large flat facets.  

\subsection{Density Function Theory Calculations}
First principles (DFT) calculations were performed with full-potential linear augmented wave formalism as implemented in WIEN2k \cite{DFTref1}. Exchange-correlation energies are calculated with Perdew-Burke-Erzenhof (PBE) functional \cite{DFTref2}. An 8x8x8 unshifted k-point grid is used in the Brillouin Zone of the primitive body-centered tetragonal cell. Results reported were double-checked by repeating calculations in Projector Augmented Wave (PAW) formalism, as implemented in VASP, and no disagreements were found \cite{DFTref3}.

{\bf Acknowledgments:} TMM acknowledges support of start-up funds from the Johns Hopkins University as well as the David and Lucile Packard Foundation. KMS acknowledges the NSF Materials Research Science and Engineering Centers (MRSEC) program Grant No. DMR-1120296, AFOSR grants FA9550-11-1-0033 and FA9550-12-1-0335, and the NSF CAREER Grant No. DMR-0847385. The dilution refrigerator used in this study was funded through the National Science Foundation Major Research Instrumentation Program, grant NSF DMR-0821005. Crystal growth supported by DOE, Office of Science, Basic Energy Sciences, Division of Materials Sciences and Engineering (The Institute for Quantum Matter, Award DE-FG02-08ER46544).

\newpage

\begin{figure}
\includegraphics[width=5.5in]{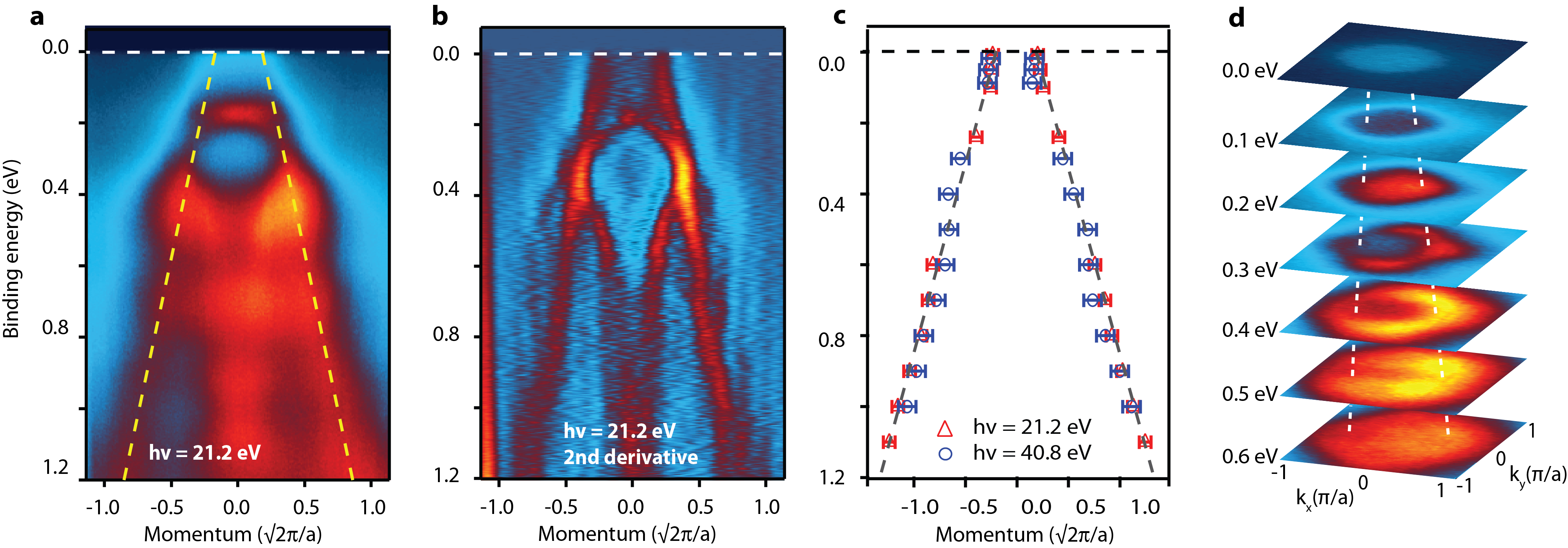}
\caption{(Color online) {\bf Dirac-like surface states in the topological superconductor Tl$_5$Te$_3$. (a)} ARPES spectrum taken with He I$\alpha$ (h$\nu$ = 21.2 eV) photons along the (0,0) -- ($\pi$,$\pi$) direction showing the coexistence of prominent bulk states and a linearly dispersing surface state (yellow dashed lines) near E$_F$. {\bf (b)} Second derivative plot of the measured experimental data in {\bf a}, where the surface state is more clearly resolved. {\bf (c)} MDC
 fits for the data taken with He I$\alpha$ and He II$\alpha$ (h$\nu$ = 21.2 eV and 40.8 eV) photons, showing no substantial dependence on photon energy. {\bf (d)} Momentum-resolved intensity maps at a series of binding energies which show the single Dirac cone feature centered at the zone center.}
\label{figTlTI}
\end{figure}

\end{document}